\documentclass[aip,amssymb,amsmath]{revtex4}
\usepackage{graphicx}
\usepackage{dcolumn}
\usepackage{subfigure}
\usepackage{rotating}

\begin{document}


\title{A waveguide-coupled thermally-isolated radiometric source}

\author{K. Rostem}
\author{D. T. Chuss}
\author{N. P. Lourie}
\author{G. M. Voellmer}
\author{E. J. Wollack}

\affiliation{NASA Goddard Space Flight Center, 8800 Greenbelt Road, Greenbelt MD 20771}
\email{karwan.rostem@nasa.gov}

\date{\today}

\begin{abstract} 

The design and validation of a dual polarization source for waveguide-coupled millimeter and sub-millimeter wave cryogenic sensors is presented. The thermal source is a waveguide mounted absorbing conical dielectric taper. The absorber is thermally isolated with a kinematic suspension that allows the guide to be heat sunk to the lowest bath temperature of the cryogenic system. This approach enables the thermal emission from the metallic waveguide walls to be subdominant to that from the source. The use of low thermal conductivity Kevlar threads for the kinematic mount effectively decouples the absorber from the sensor cold stage. Hence, the absorber can be heated to significantly higher temperatures than the sensor with negligible conductive loading. The kinematic suspension provides high mechanical repeatability and reliability with thermal cycling. A 33-50 GHz blackbody source demonstrates an emissivity of 0.999 over the full waveguide band where the dominant deviation from unity arrises from the waveguide ohmic loss. The observed thermal time constant of the source is 40 s when the absorber temperature is 15 K. The specific heat of the lossy dielectric MF-117 is well approximated by $C_v(T)=0.12\,T\,^{2.06}$ mJ g$^{-1}$ K$^{-1}$ between 3.5 K and 15 K.
 
\end{abstract}

\maketitle

\section{Introduction}

Accurate radiometric metrology is essential for the successful characterization of cryogenic sensors. At millimeter and sub-millimeter wavelengths, a thermal blackbody source of known temperature can be used as a radiometric source~\cite{Addario, Peterson, Grath, Mather, Pisano, Vizard, Wollack, Alison}. The source can function as a stable reference for the characterization of the sensor stability or as an absolute power standard if the calibrator's emissivity, thermal distribution, and radiometric power coupling are understood in detail. 

In single-mode waveguide applications, tapered lossy dielectric absorbers in the shape of vanes~\cite{Addario, Grath}, cones, or wedges~\cite{Alison, Wollack} are commonly used. The absorber is typically in contact with the waveguide wall and thermal isolation from the sensor is achieved through an intervening section of low thermal conductivity metallic waveguide constructed from brass, bronze, or stainless steel. Radiometric emission and thermal conduction resulting from the use of a waveguide thermal break can significantly complicate the use and achievable accuracy for cryogenic sensor characterization. Accurate metrology then requires a detailed knowledge of temperature gradients throughout the calibrator. This greatly increases the uncertainty in the flux and complexity of the source calibration.  Heating such a configuration not only induces significant temperature gradients in the source but can also impart mechanical stresses and dimensional changes that limit reliability and radiometric accuracy. Alternatively, a thermally isolated thin-film absorber can be inserted through a thin slot in the waveguide wall~\cite{Addario, Grath}; however, the absorbing vane position, and thus the source reflectivity, can potentially vary with temperature and display microphonic sensitivity.

\begin{figure}[htbp]
\begin{center}
\includegraphics[width=8cm]{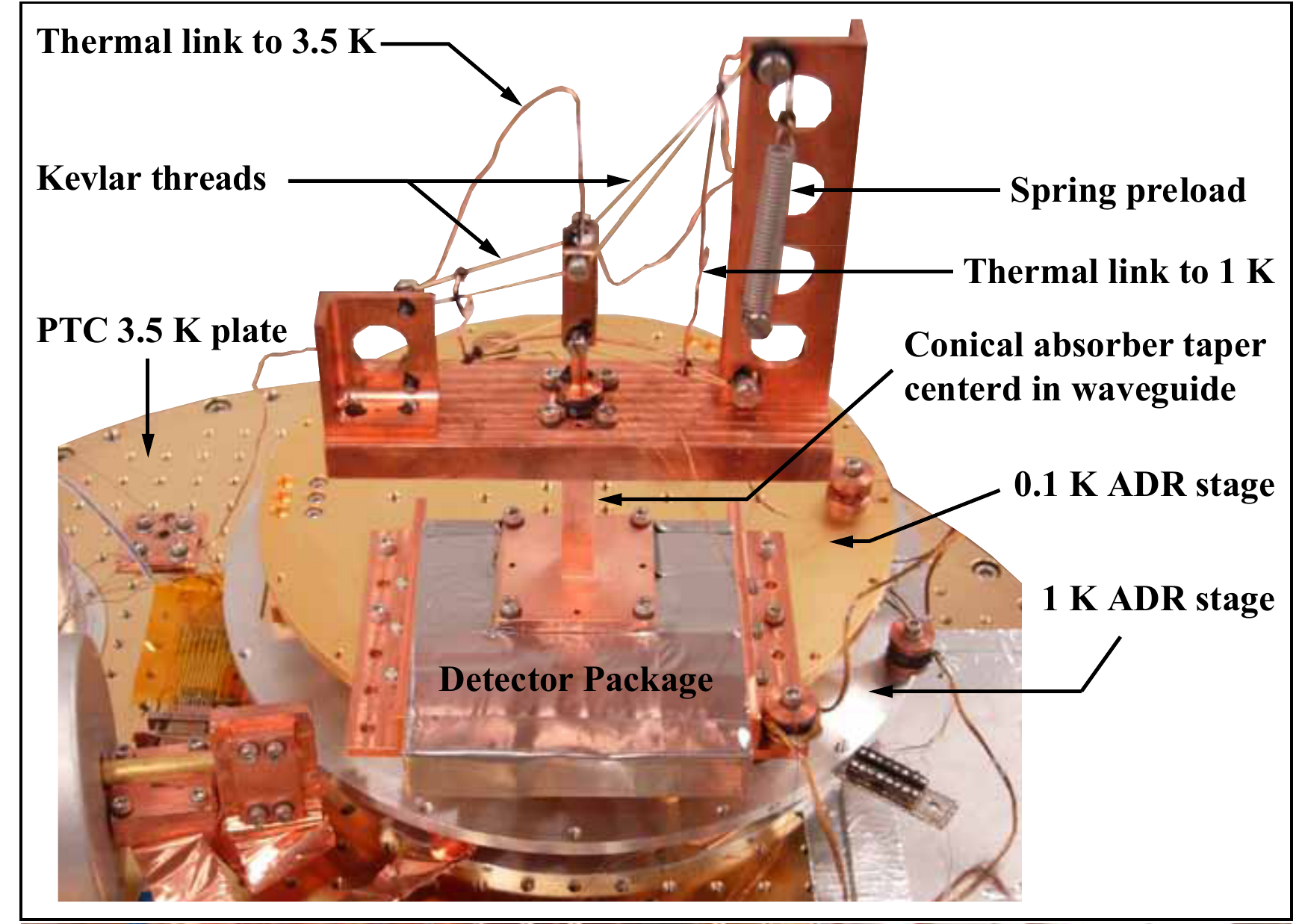}
\caption{\label{fig:actual} Photograph of the source in a cryogen-free refrigeration system. The tapered absorber is centered within and thermally isolated from a square electro-formed copper waveguide using a kinematic suspension. }
\end{center}
\end{figure}

For an absolute power standard, it is desirable to have a uniform temperature distribution throughout the absorber geometry. The flux is then characterized by a single physical temperature that can be varied via Joule heating of a resistor in intimate thermal contact with the absorber. In this paper, we present a dual polarization radiometric source that approximates this goal and is well suited for the characterization of waveguide-coupled sensors at sub-Kelvin temperatures. Figure~\ref{fig:actual} shows the source attached to a detector package cooled to 0.1 K. The kinematic suspension of the absorber has the following advantages: (i) The absorber is conductively decoupled from the sensor environment. (ii) The position of the absorber is invariant relative to the guide walls, thus minimizing changes in reflectance with temperature. (iii) The metallic walls of the mount have negligible in-band radiometric emission relative to the waveguide absorptive termination. 

In Sec.~\ref{sec:mechanics}, we describe the elements, mechanical design, and assembly of the source. In Sec.~\ref{sec:therm}, we report measurements of the thermal response of the source as tested in the cryogenic setup shown in Fig.~\ref{fig:actual}. In Sec.~\ref{sec:taper-design}, we report on the measured electromagnetic performance of the tapered absorber. Analysis of parameter tolerances required and  achieved for the source is presented in each section. 

\section{Source Design and Assembly\label{sec:mechanics}}

\begin{figure}[htbp]
\begin{center}
\includegraphics[width=8cm]{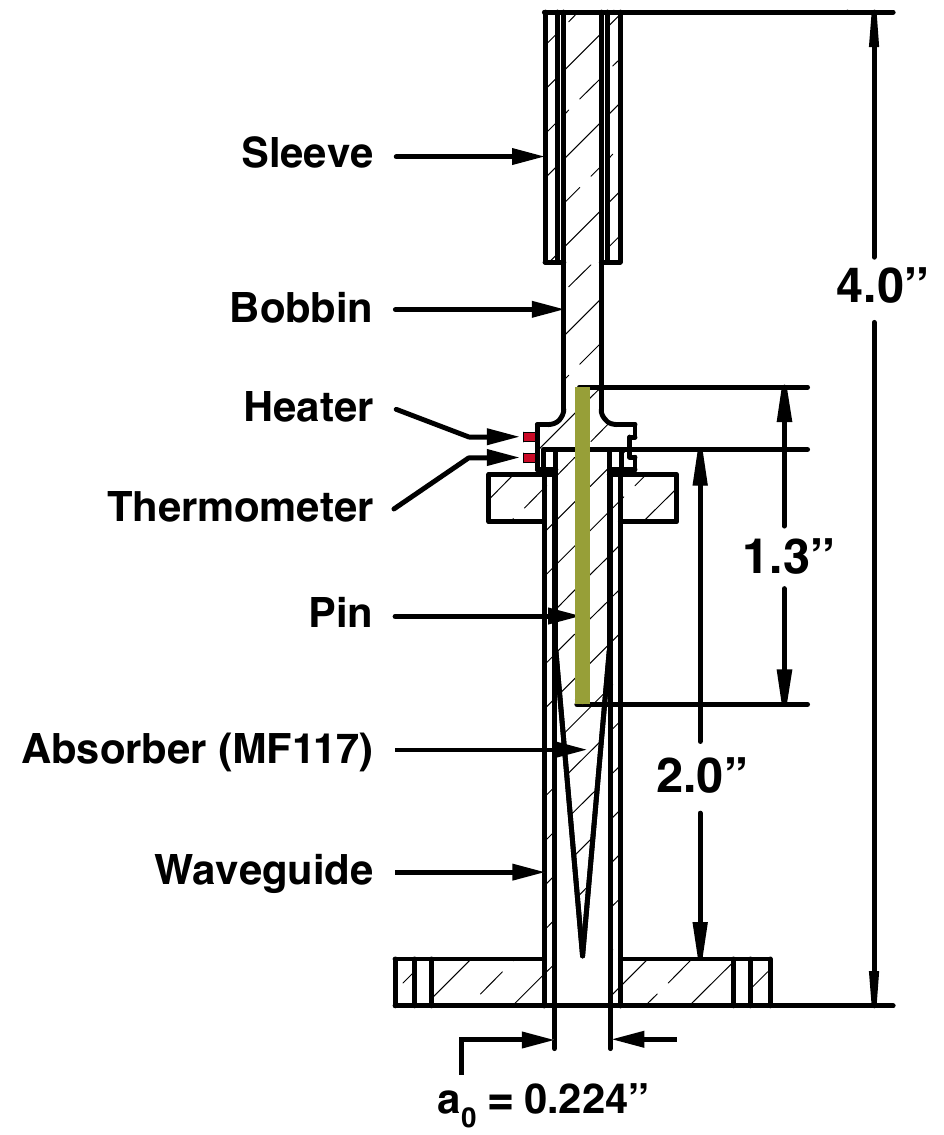}
\caption{\label{fig:assembly} (Color online) Sketch of the source. The absorber is bonded to a copper pin (colored green) that reaches a depth of 1'' into the absorber to effect isothermal response. The bobbin-absorber is inserted into a sleeve, centered within the waveguide with shims, and bonded to the sleeve. }
\end{center}
\end{figure}

The source is comprised of a lossy conical taper centered in a waveguide, as shown in Fig~\ref{fig:assembly}. The absorbing cone is made from Eccosorb MF-117~\cite{mf117}, a commercially available machinable lossy dielectric realized from an epoxy loaded with a conductive powder filler. To achieve a uniform thermal distribution, the absorber is bonded to an annealed copper pin that is inserted 1'' into the absorber. This assembly is then bonded to a copper mounting bobbin with conductive silver epoxy Epotek H20E~\cite{H20E}. A  calibrated LakeShore Cernox thermometer (CX-1050) traceable to NIST and a 1 k$\Omega$ surface mount thin-film metal resistor that functions as a heater are bonded to the bobbin with Stycast 2850FT~\cite{Stycast2850FT}. We use 0.003'' diameter twisted-pair NbTi leads to make electrical contact with the thermometer and heater. Approximately 2'' of each twisted pair is bonded to the bobbin for thermal anchoring, and 5'' of each twisted pair serves as a thermal break from a 3.5 K cold stage.   

\begin{figure}[htbp]
\begin{center}
\includegraphics[width=8cm]{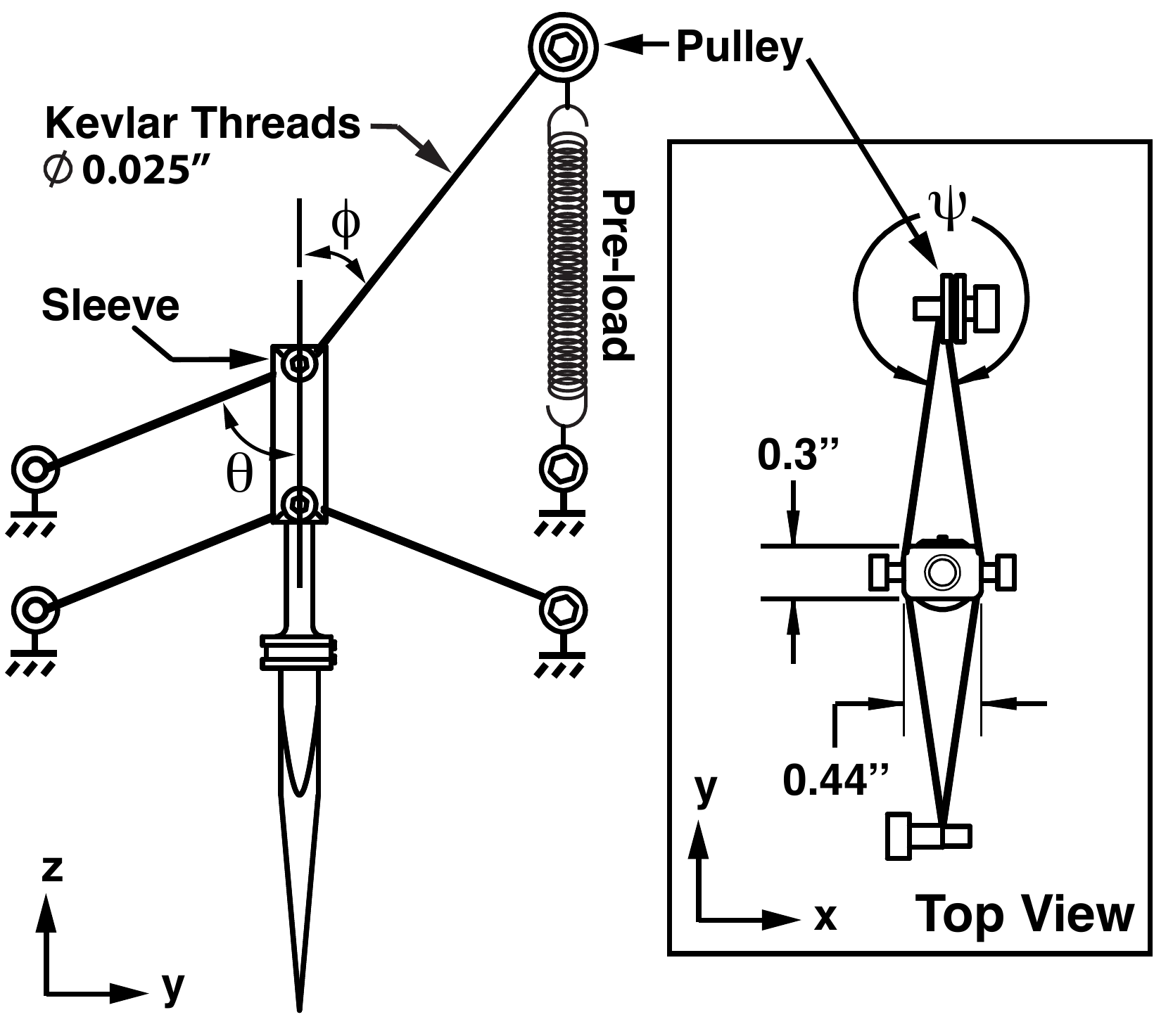}
\caption{\label{fig:kinematic}Detailed sketch of the kinematic suspension. The spring preload ensures tension is maintained throughout thermal cycling. The magnitude of the preload (60 N) is at 20\% of the tensile strength of Kevlar. The kinematic suspension is rigid in the x-y plane, and does not allow rotation about the z axis. The only degree of freedom is translation in the z direction, which is less than 0.5 mm. Approximate dimensions of the angles are as follows: $\phi=30^\circ$, $\theta=60^\circ$, and $\psi=340^\circ$. }
\end{center}
\end{figure}

The bobbin is bonded to a copper sleeve that is held kinematically with an aramid fiber, Kevlar~\cite{dupont}, suspension as shown in Fig.~\ref{fig:kinematic}. The sleeve is under a tensile preload applied with a steel spring. The preload is maintained throughout a thermal cycle. As a general rule, bends and knots decrease the tensile strength of Kevlar threads by $\sim$50\%. We applied a preload of 60 N, $\sim$ 20\% of the Kevlar tensile strength, to ensure adequate margin.  

The kinematic suspension geometry ensures the position of the absorber remains invariant with respect to the waveguide walls when the source is cooled to cryogenic temperatures. The only unconstrained degree of freedom of the absorber is that associated with motion in the vertical direction (z axis in Fig~\ref{fig:kinematic}), which is less than 0.5 mm. Given the absorber's reflectivity, this motion has negligible influence on electromagnetic performance of the source in practice. 

Prior to bonding the bobbin to the sleeve, an electroformed square copper waveguide is attached to the source and aligned to the absorber with 125 $\mu$m plastic shims~\cite{plastic}. The separation between the absorber and the waveguide walls is less than 250 $\mu$m. The static centering tolerance, which limits higher order mode conversion in the waveguide, is 25 $\mu$m on each side and is acceptable for the frequency range of interest (33-50 GHz). The required centering tolerance scales directly with wavelength and a few microns can be achieved with this methodology, thus enabling implementation at higher frequencies. 
 
\section{Thermal Response\label{sec:therm}}

\begin{figure}[htbp]
\begin{center}
\includegraphics[scale=1]{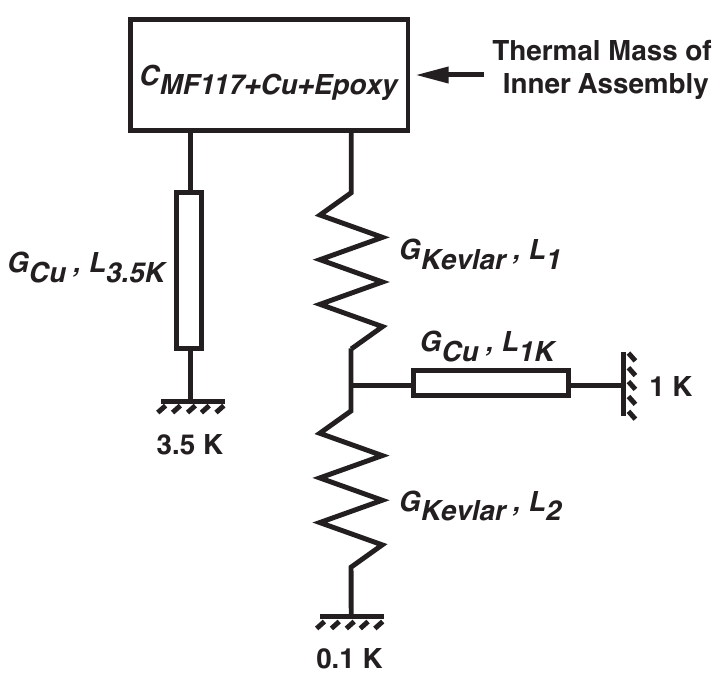}
\caption{\label{fig:therm-circ}Thermal circuit of the source when installed in a cryogenic system. The heat capacity of the source is represented by a lumped element that is dominated by the lossy dielectric MF-117 absorber. The thermal conductivity of the copper wires that attach to the sleeve is much larger than the Kevlar threads. The source lengths $L_{3.5K}=26$ cm and $L_{1K}=6$ cm can be tailored to meet the desired experimental thermal time constant.  }
\end{center}
\end{figure}

In our cryostat, the source can be thermally anchored to one of three thermal baths: the first stage of an Adiabatic Demagnetization Refrigerator (ADR) at 0.1 K with a cooling capacity of 100 mJ, the second stage of the ADR at 1 K with a cooling capacity of 1 J, and a 3.5 K plate that is cooled at a rate of 278 mW with a Pulse-Tube Cooler (PTC). Figure~\ref{fig:therm-circ} shows the thermal circuit for the source used in this environment. The implementation satisfies the following requirements: (i) For characterization of astronomical detectors operating at millimeter wavelengths, the absorber must reach $\sim$10-20 K in order to simulate the sky emissivity for a typical ground-based observatory~\cite{Pardo}. For radiometric flux measurements, the thermalization time-constant should be short compared to timescales over which the sensor properties remain invariant, or the so-called device $1/f$-knee. The source should also be capable of regulation at a constant temperature for use as a stabilized load. (ii) The heat transfer to the 1 K and 0.1 K stages of the ADR must be minimal and not significantly impact refrigerator performance. This constraint is set by the cooling capacities of the ADR salt pills. (iii) The PTC must handle the heat transfer from the absorber when heated to its maximum temperature. 

The source is thermally connected to the 3.5 K thermal bath with commercially available oxygen-free (99.9\% purity) solid copper wire. The diameter of the wire is 0.032''; its length is used as a free parameter to achieve the desired thermal conduction. Since the conductivity of Kevlar is very low compared to copper (approximately a factor of 10$^4$ difference below 15 K~\cite{NIST-cryogenics}), the electrical heat dissipated in the heater is effectively conducted to the 3.5 K bath. Additional copper wires thermally anchor each Kevlar thread to the 1 K ADR stage at a location approximately half way between the 0.1 K and 3.5 K ends. This limits the heat flow from the source to the 0.1 K ADR stage that provides cooling for the sensor under test. The length $L_{1K}$ of these wires is kept as short as possible to achieve the highest thermal conduction. In the implemented configuration, the dissipated power on the 0.1 K ADR stage is 0.5 $\mu W$.

\begin{figure}[htbp]
\begin{center}
\includegraphics[width=8cm]{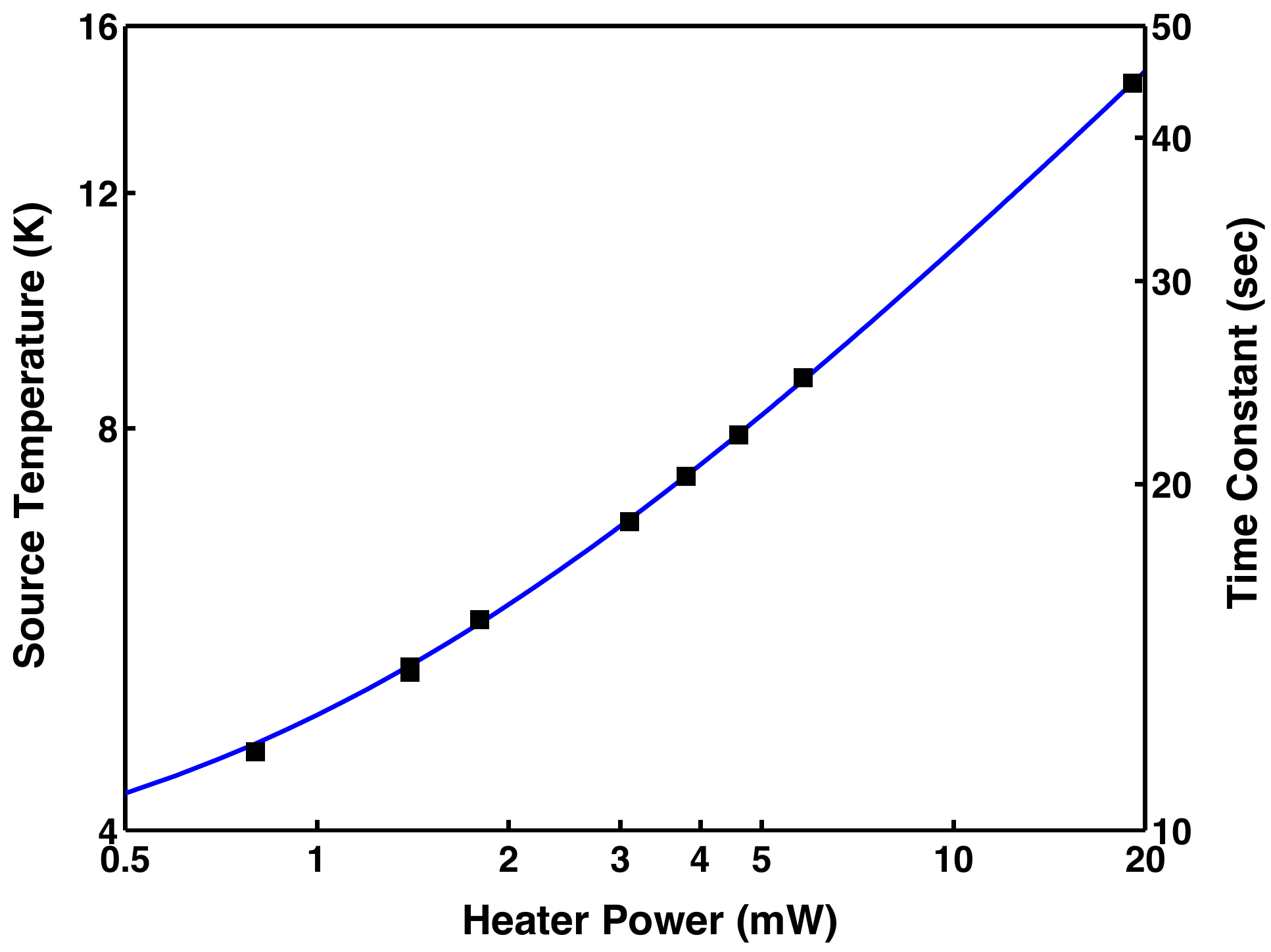}
\caption{\label{fig:tau-power}Measured dependence of the source temperature and time constant on heater power (filled squares). The solid line is a power law fit to the data of the form $P=K(T^n-T_B^n)$, where $K=0.072$ mW K$^{-n}$ and $n=2.11$ are the fit parameters, $P$ is the heater power, $T$ is the source temperature, and $T_B=3.5$ K is the bath temperature.}
\end{center}
\end{figure}

The thermal time constant of the absorber is 40 s when heated to 15 K. A time constant of $\sim10$ minutes was also observed with an amplitude (0.2\%) significantly smaller than that of the principal thermal response time. In the analysis that follows, we focus on the dominant mode for thermal equilibration as it can be clearly associated with the thermal mass of the absorber. Figure~\ref{fig:tau-power} shows the relationship between the measured source temperature, thermal time constant and applied heater power. 

The thermal conductivity of the copper wire and specific heat capacity of the MF-117 absorber are essential parameters in the thermal model of the source. Although these parameters were initially estimated from results found in the literature~\cite{Peterson}, {\it in-situ} measurements of the thermal response enabled a direct determination. 

From the source temperature and heater power functional form shown in Fig.~\ref{fig:tau-power}, the thermal conductivity of the copper wire was determined and found to be comparable to the conductivity of OFHC copper with a resistance-residual-ratio (RRR) of $\sim100$. This RRR is also consistent with measurements of the electrical resistivity of copper wire from the same spool. 

\begin{figure}[htbp]
\begin{center}
\includegraphics[width=8cm]{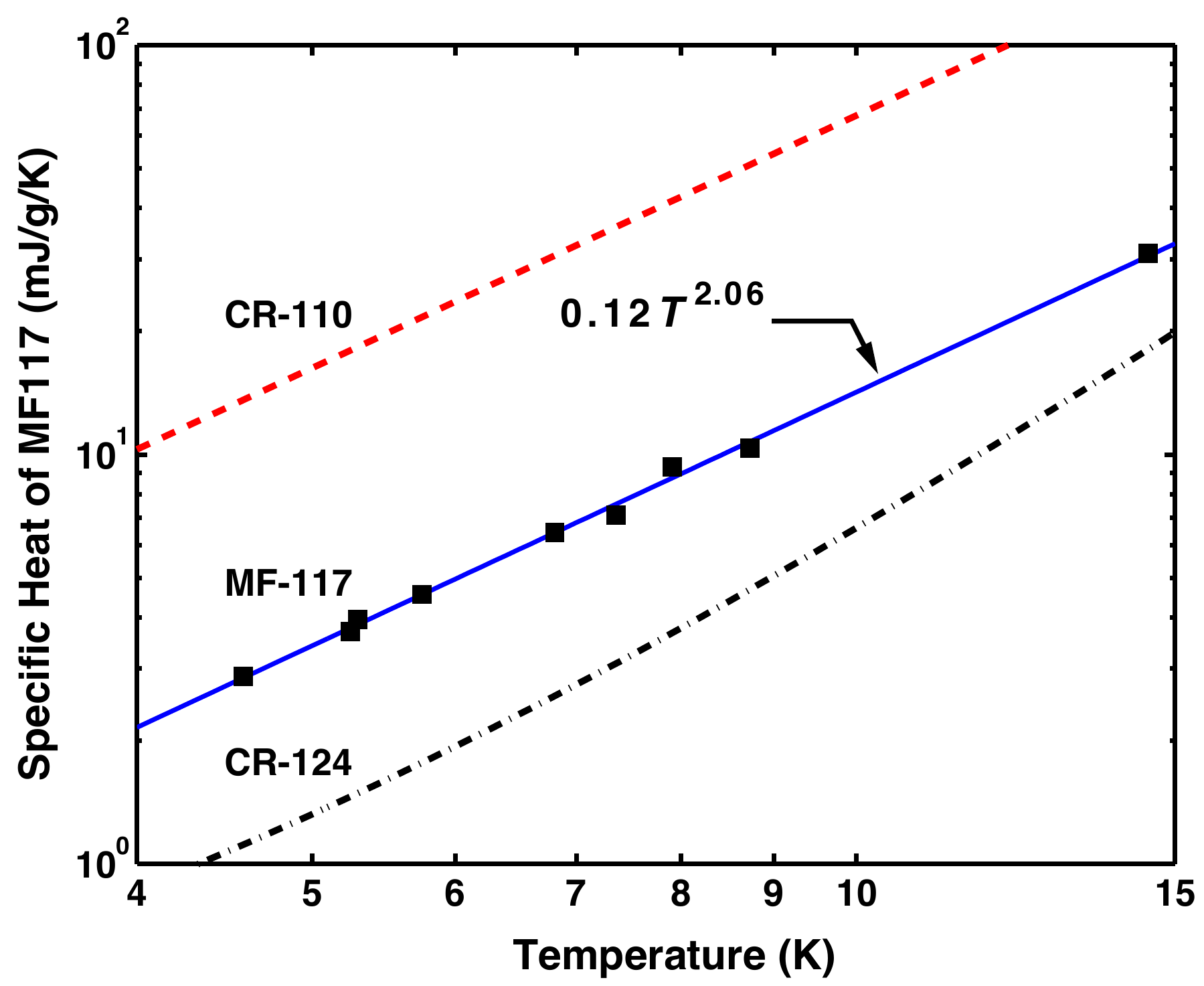}
\caption{\label{fig:MF117-Cv}(Color Online) Measurements of the specific heat of Eccosorb MF-117. The solid line is a fit to the data (filled squares). The specific heat of Eccosorb CR-110 and CR-124 are from Ref.~\cite{Peterson}~and~\cite{Wikus} respectively.}
\end{center}
\end{figure}

The total heat capacity of the source is $C(T) =\tau(T)\cdot G(T)$, where $\tau(T)$ is the measured time constant relation shown in Fig.~\ref{fig:tau-power}, and $G(T)$ is the thermal conductance of the copper wire $L_{3.5K}$ in Fig.~\ref{fig:therm-circ}. Taking into account the specific heat contributions from the copper~\cite{Kittel}, H20E~\cite{Weyhe}, and Stycast 2850FT~\cite{Javorsk} in the inner-assembly, and the mass of the absorber (3.26 g), we estimate the specific heat of MF-117 with the functional form $C_v(T)=0.12\,T\,^{2.06}$ mJ g$^{-1}$ K$^{-1}$ between 3.5 K and 15 K. The last two digits used in the naming convention of the MF stock are correlated with the metallic volume filling fraction of the mixture. Lacking a published value for the $C_v(T)$ of MF-117, we compare the measured response to CR-110~\cite{Peterson} and CR-124~\cite{Wikus} as shown in Fig.~\ref{fig:MF117-Cv} which bound the filling fraction of the formulation in use. The CR series of lossy dielectric is available as a casting resin~\cite{mf117} and when mixed and cured properly, is identical to the machinable MF stock. 

For amorphous materials in general, the specific heat at low temperatures ($<10$ K) can be described by the function $C_v(T)=aT + bT^3$, where the linear term is associated with excess heat capacity common to amorphous materials~\cite{Zeller,Kelham}, and the cubic term is the Debye contribution. The data in Ref.~\cite{Wikus} is approximated by this form, although the linear term may be associated with the high metallic fraction in CR-124. Ref.~\cite{Peterson} and this paper report a $T^2$ relation for CR-110 and MF-117. The relative values of the specific heat of the materials shown in Fig.~\ref{fig:MF117-Cv} can be understood in terms of the volume filling fraction of metal in the dielectric mixtures. At higher filling fractions, the metal is displacing the high specific heat amorphous dielectric. In general, additional measurements of loaded epoxy resins are required to gain further insight into how the thermal properties change as a function of load concentration and preparation method.  

For our application, the use of a lossy dielectric absorber is acceptable from an electromagnetic and thermal performance standard. However, if time constants of less than a second are required, an appropriately shaped crystalline quartz or silicon wedge coated with a thin-film metallization can be used as an absorber~\cite{Addario, Grath} with the kinematic suspension demonstrated here.

\section{Electromagnetic Performance\label{sec:taper-design}}
\subsection{Reflectance of a lossy conical taper}

\begin{figure}[htbp]
\begin{center}
\includegraphics[width=8cm]{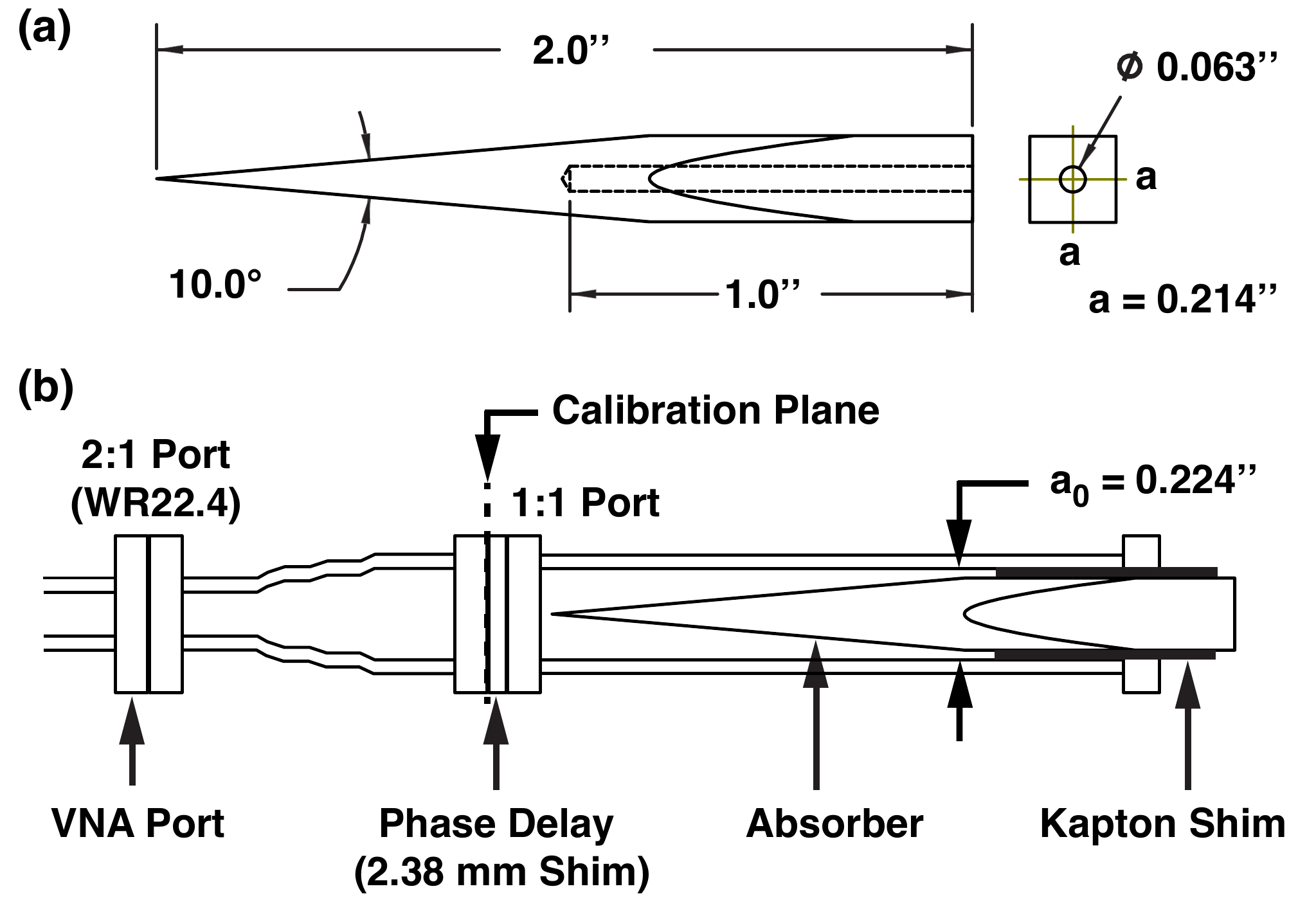}
\caption{\label{fig:waveguide-assembly}(a) Sketch of the conical taper that functions as a lossy adiabatic impedance transformer when placed in a square waveguide housing. (b) Sketch of the waveguide assembly employed for the power reflection measurement of the absorber. The reflectance is measured with and without a quarter-wave delay shim inserted at the calibration plane. }
\end{center}
\end{figure}

Figure~\ref{fig:waveguide-assembly}(a) shows a detailed diagram of the conical taper. The use of MF-117 was largely determined by the availability of an existing absorptive conical taper. In practice, other conductively loaded dielectrics can be used if desired~\cite{Wollack-SC}. Our experience shows that the electromagnetic properties of MF-117 can be approximated by a complex relative dielectric permittivity $\epsilon_r \simeq 8+i1$ and magnetic loss $\mu_r\approx1$~\cite{Wollack-SC}. At 33 GHz, the penetration depth in a bulk sample of the material is $\delta \simeq \lambda_o/(\pi \sqrt{2 \Im[\epsilon_r]}) \approx 2$ mm, where $\lambda_o$ is the radiation wavelength in free space. The attenuation per unit length for bulk samples of this class of lossy dielectrics changes by $\sim$10\% between 293 K and 5 K~\cite{Kerr, Wollack-SC} as a result of the use of disordered metal alloys as the loading media. For the adiabatic cone geometry employed, the reflectance and emittance are essentially independent of $\epsilon_r$ (see Fig.~1 in Ref.~\cite{Wollack}). As a result of this choice, the radiometric performance of the MF-117 absorber is insensitive to temperature. 

The reflectance of the absorber assembly was measured at room temperature using an Agilent Technologies N5245A Programmable Network Analyzer calibrated in WR22.4 with Thru-Reflect-Line (TRL) calibration. Figure~\ref{fig:waveguide-assembly}(b) is a sketch of the measurement assembly. The absorber is inserted into an electroformed square copper waveguide and centered within the guide with 125 $\mu$m Kapton~\cite{dupont} shims. The broadwall width of the guide is $a_0 = 0.224$'' for compatibility with existing transitions and calibration components in WR22.4 rectangular waveguide. The guide length is set by the cone length, such that the tip of the cone does not extend beyond the guide port when the absorber is fully inserted. The source's complex reflection amplitude was measured as function of frequency with ($\Gamma_1$) and without ($\Gamma_2$) a quarter-wave delay shim inserted at the calibration plane shown in Fig.~\ref{fig:waveguide-assembly}(b)\cite{Eimer}. The absorber's complex reflection amplitude is given by,
\begin{eqnarray}
\Gamma_{abs} = \frac{\Gamma_1 - \Gamma_2}{1-\exp(i\phi)},
\label{eqn:cal}
\end{eqnarray}
where $\phi=4\pi L/\lambda_g$ is the phase shift associated with the shim thickness $L$, $\lambda_g = \lambda_o/\sqrt{1-(\lambda_o/\lambda_c)^2}$ is the guide wavelength, and $\lambda_c=2a_0$ is the guide cut-off wavelength. 

\begin{figure}[htbp]
\begin{center}
\includegraphics[width=8cm]{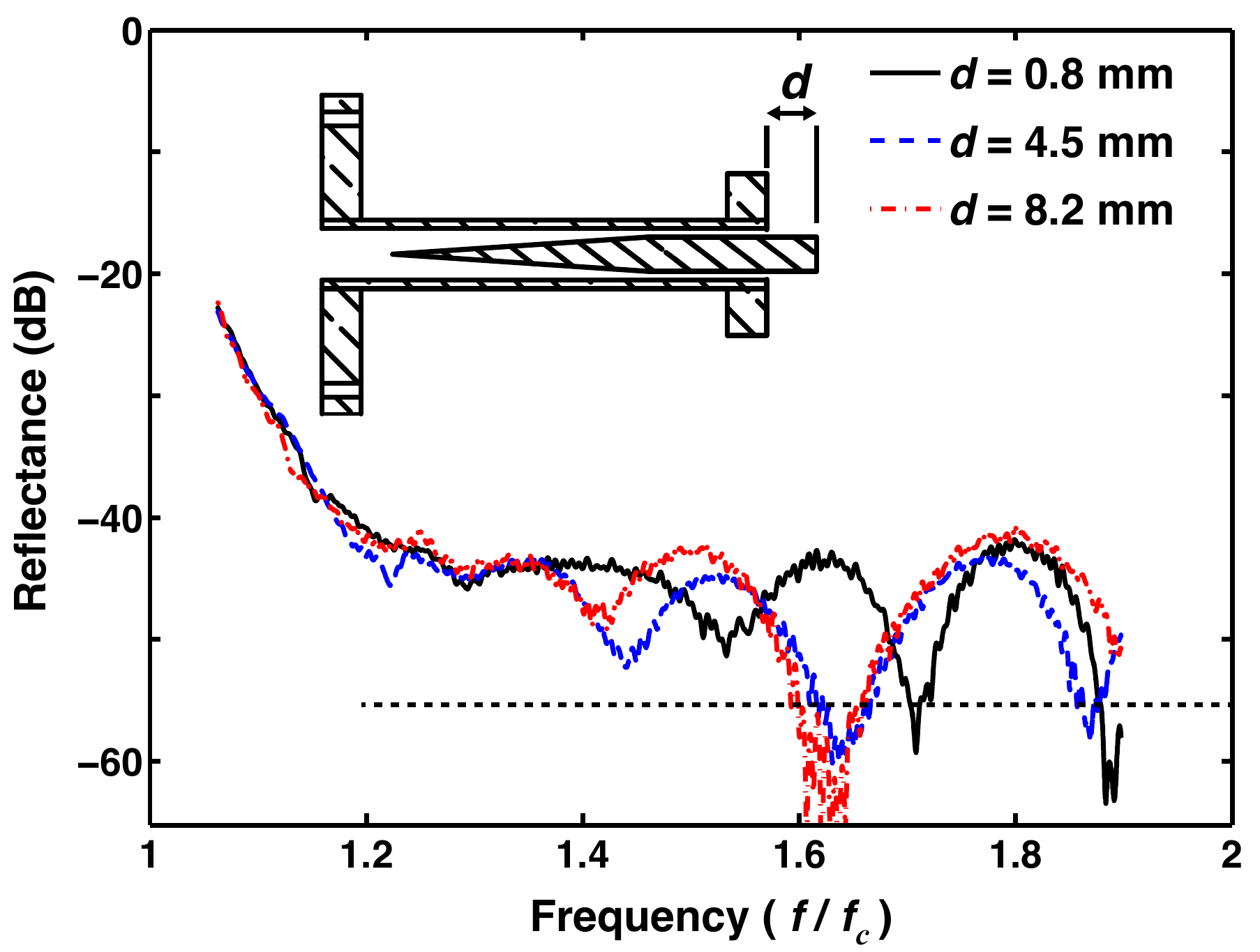} 
\caption{\label{fig:cone-reflectance}(Color online) Room temperature measurements of the reflected power from the absorbing cone shown in Fig.~\ref{fig:waveguide-assembly}. The waveguide cut-off frequency is $f_c = 26.35$ GHz. The residual error of the calibration in the useful waveguide band is given by the black dashed line. The reflectance is experimentally observed and characterized by varying the absorber insertion position, $d$.}
\end{center}
\end{figure}

The upper bound on the reflectance of a lossy taper can be estimated from $A_{tip}/4b^2$, where $A_{tip}$ is the tip area of the cone and $b$ is the waveguide height~\cite{Alison}. For a tip diameter of 275 $\mu$m, a reflectance of $<-33$ dB is expected and is observed at the level of -40 dB. The reflectance as a function of the absorber insertion distance within the waveguide are shown in Fig.~\ref{fig:cone-reflectance}. The residuals of the two-tiered calibration for the network analyzer are $\sim$ -55 dB.

The electroformed copper waveguide sections are centered on the square mounting flanges and these interfaces were aligned by clamping outer flange surfaces in a machinist vise. The measured flange reflection was observed to be below -45 dB. The flanges also have 0.0635'' diameter pins that allow a maximum misalignment of 0.002''. If these pins were used for alignment in blind mate of the flanges, the reflection would be degraded to $\Delta R<-35$ dB~\cite{Hunter}. In attaching the source to the device under test with this interface, the variability in the flange reflection dominants the repeatability error. With the load in place, we observe a transmittance of $<-55$ dB through the absorber mount. 

When used as a cryogenic source, there is an uncertainty in the emitted flux that is associated with knowledge of the absorber temperature, which can arise from temperature gradients across the absorber and measurement errors in the thermometry. The observed thermal response can be modeled as a lumped heat capacity and the absorber is to a good approximation isothermal. A one-dimensional finite-element model of the absorber tapered region indicates that the tip temperature, $T_{\rm tip}$, is within 300 $\mu$K (1 $\mu$K) of the bobbin temperature when operated at 15 K (3 K). The temperature gradients are largely concentrated near the tip, at $<$5\% of the taper length. For a tapered termination centered in a guide, the power absorption mostly occurs within $\lambda_g$ from the tip~\cite{Wollack}. The uncertainty in the absorber temperature is then dominated by the thermometry. For the calibrated LakeShore Cernox thermometer bonded to the source, the uncertainty is $\pm$4 mK at 4 K and $\pm$8 mK at 15 K. Relative to the thermal conductances in the elements of the source, the boundary resistance between the thermometer and the source is negligible. 

The emitted flux at the waveguide flange is reduced by ohmic loss in the guide. Assuming a RRR of 7 for the electroformed copper, the total waveguide conduction loss within the emitting region is $\Delta \alpha < 0.001$ across the signal band. We do not correct for this term and it effectively determines the bias in the source flux at the input flange. Hence, the source is treated as a blackbody with an emissivity $0.999$ at 33-50 GHz. Table~\ref{tbl:summary} summarizes the terms contributing to the radiometric accuracy for measurement timescales large compared to the thermal time constant. 

\begin{table}[htdp]
\caption{\label{tbl:summary}Summary of source flux errors.}
\begin{center}
\begin{tabular}{l|c}
\hline
\hline
Waveguide loss bias, $\Delta \alpha$ & \, $<1\times$10$^{-3}$ \\
Thermometry uncertainty at 15 K, $\Delta T/T$ & \, 4$\times$10$^{-4}$ \\
Flange reflectance variability, $\Delta R$ & \, $<3\times$10$^{-4}$ \\
Taper emissivity uncertainty, $\Delta \epsilon$ & \, 1$\times$10$^{-4}$ \\
Absorber gradient uncertainty at 15 K, $\Delta T_{\rm tip}/T$\,   & \, $\sim 2\times$10$^{-5}$ \\
\hline
\end{tabular}
\end{center}
\end{table}%

\subsection{Available Source Power}
 
The power emitted by the source can be calculated by treating the absorber as a blackbody radiator. However, the power observed at the input flange is reduced by reflection due to flange misalignment. The available source single-mode thermal power as a function of temperature is given by
\begin{eqnarray}
P_s(T) = \int_{0}^{\infty} \left(1-R(\nu) \right) \, \epsilon(\nu) \, h\nu\, F(\nu)\,n(T,\nu)\,d\nu,
\label{eqn:power}
\end{eqnarray}
where $T$ is the source temperature, $\epsilon(\nu)$ is the emissivity of the source, $n(T,\nu)=1/[\exp(h\nu/k_BT)-1]$ is the photon Bose-Einstein distribution, $h$ is the Planck constant, and $k_B$ is the Boltzmann constant. The function $F(\nu)$ describes the effective band definition filters used to limit the power of the source. $R(\nu)$ accounts for the power reflection due to the flange misalignment. $R(\nu)$ is negligible and $\epsilon(\nu)$ is nearly unity over the full waveguide band as shown in Table~\ref{tbl:summary}.

In the Rayleigh-Jeans limit $h\nu/k_BT\ll1$, Eq.~\ref{eqn:power} reduces to $P_s \approx  kT \int_{0}^{\infty} F(\nu)\,d\nu$. The integral over $F(\nu)$ is equivalent to an effective radiometric signal bandwidth, and the available power can be written in the familiar form $P_s \cong  k_BT\Delta\nu_{RF}$. However, to accurately describe the available power for the source, Eq.~\ref{eqn:power} should be used in general~\cite{Jonas} since $F(\nu)$ can extend into the Wien limit $h\nu/k_BT>1$, where $h\nu\, n(T,\nu)$ decays exponentially. 

\begin{figure}[htbp]
\begin{center}
\includegraphics[width=8cm]{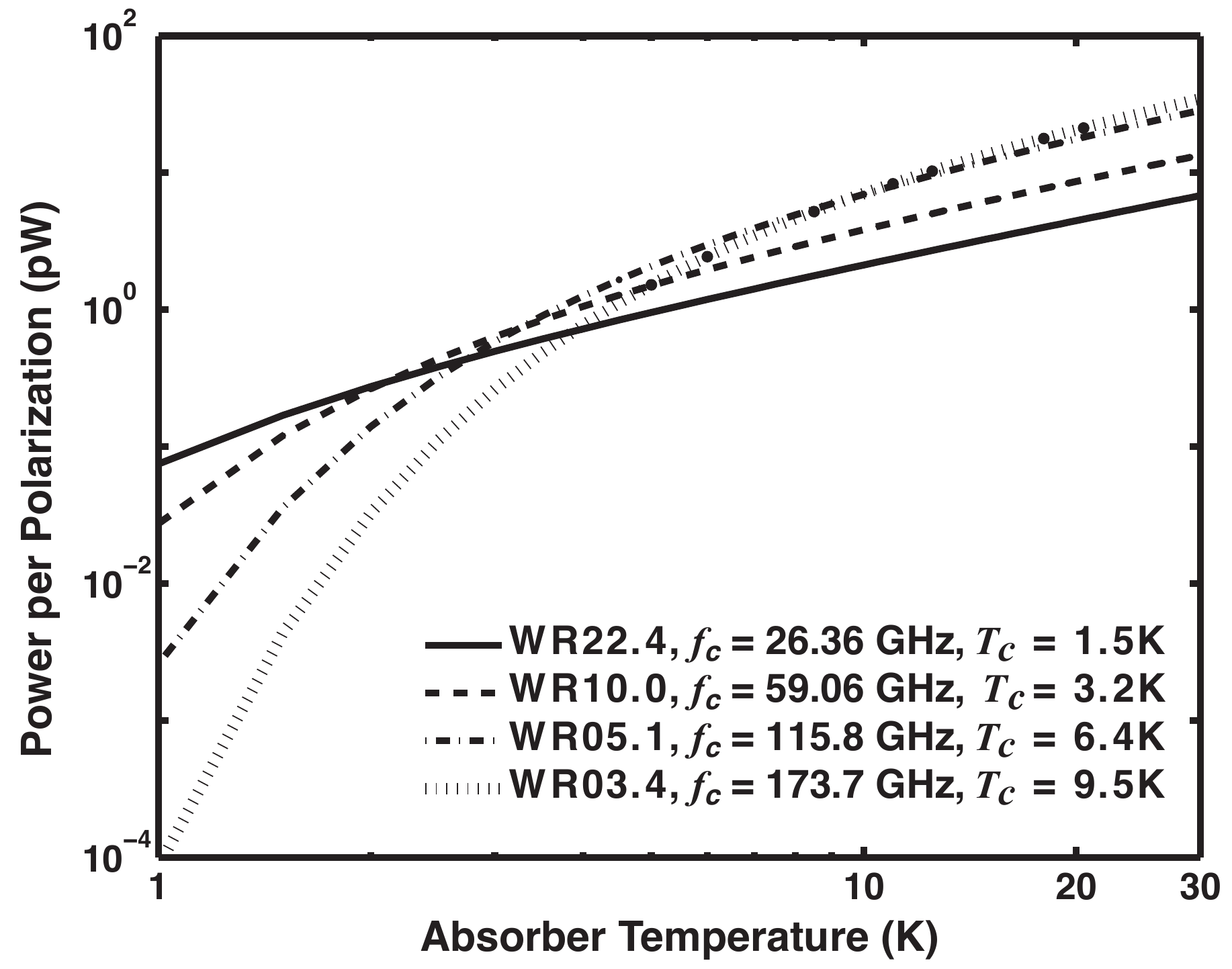}
\caption{\label{fig:powerPerPol}Available single-mode power from the thermal source in frequency bands defined by single-mode rectangular waveguide between $1.3f_c$ and $1.9f_c$. The available power drops rapidly below a cut-off temperature $T_c = 12f_c h/\pi^2k_B$. The coupling loss due to flange reflection has been set to zero. }
\end{center}
\end{figure}

A waveguide is a high-pass filter structure and its cut-off frequency, $f_c$, specifies the minimum source frequency that can propagate. The maximum frequency at which the source emits is not bound and tends to infinity with increasing temperature. However, as the source temperature decreases, the equivalent frequency of the exponential decay in $h\nu\, n(T,\nu)$ can become less than the upper band edge frequency $2f_c$ where the source would find use for dual-polarization illumination of a device. The available power drops rapidly below a cut-off temperature that can be defined as $T_c = 12f_c h/\pi^2k_B$.  An estimate of the available power for selected frequency bands is shown in Fig.~\ref{fig:powerPerPol} and indicates this general behavior. A low-pass filter whose propagation characteristics have been independently characterized can be employed to define the upper band edge at $2f_c$. In this case, the uncertainty in $F(\nu)$ dominates the absolute calibration error. Alternatively, for integrated bolometric waveguide sensors, the band definition filters are typically an element of the sensor under test~\cite{Rostem} and suitable determination of the frequency response is required to specify the available power to the detector.

For higher frequency bands ($>100$ GHz), smaller conical tapers can be constructed that will have the additional advantage of a lower thermal mass than the taper described here. An alternative approach is to use an absorber designed for low frequency operation in conjunction with a waveguide transition that adiabatically couples the relevant modes to a higher frequency device. Absorber tip diameters of 25 $\mu$m have been realized for use as waveguide terminations and can be used in scaling the source for operation at shorter wavelengths. In this regime, flange alignment presents a greater challenge and can be addressed through the use of ring center flanges~\cite{Kerr-Ring} to maintain the required tolerances.

\section{Conclusion}

A dual polarization thermal source for millimeter and sub-millimeter waveguide-coupled sensors has been presented. The design principles outlined are directly applicable to improving the performance of single polarization waveguide calibrators. The source can be used as an absolute power reference for metrology and as a stable source for characterization of sensor stability. The thermal source has an absorptive dielectric taper centered in a waveguide. The source is thermally isolated from the waveguide with a kinematic suspension. The thermal response time is 40 s at 15 K, which is desirable for the characterization of sensor stability and optical efficiencies below this temperature. The design is mechanically robust, can be scaled to different wavebands, and can be readily reconfigured for other thermal environments and radiometric demands. 
\clearpage
\section{Acknowledgement}

We gratefully acknowledge financial support from the NASA ROSES/APRA program. K. Rostem was supported by an appointment to the NASA Postdoctoral Program at Goddard Space Flight Center. We thank Paul Cursey for fabrication and metrology support. 

\clearpage
\bibliography{report}

\end{document}